\documentclass[twocolumn]{aastex62}

\usepackage{amsmath}

\usepackage{xcolor}
\definecolor{navyblue}{rgb}{0.0, 0.0, 0.5}
\definecolor{royalblue}{rgb}{0.25, 0.41, 0.88}
\definecolor{cadmiumgreen}{rgb}{0.0, 0.42, 0.24}
\definecolor{blue-violet}{rgb}{0.54, 0.17, 0.89}
\definecolor{darkviolet}{rgb}{0.58, 0.0, 0.83}
\definecolor{magenta}{rgb}{1.0, 0.0, 0.56}
\definecolor{darkspringgreen}{rgb}{0.09, 0.45, 0.27}
\definecolor{royalblue}{rgb}{0.25, 0.41, 0.88}

\newcommand{\daniel}[1]{{ #1}}
\newcommand{\dan}[1]{\textcolor{magenta}{}}
\newcommand{\rr}[1]{{ #1}}
\newcommand{\maya}[1]{{ #1}}
\newcommand{\holz}[1]{{ #1}}
\graphicspath{{./}{figures/}}
\begin{document}

\title{A standard siren measurement of the Hubble constant from GW170817 without the electromagnetic counterpart
}

\author{M.~Fishbach}
\affiliation{Department of Astronomy and Astrophysics, University of Chicago, Chicago, IL 60637, USA}

\author{R.~Gray}
\affiliation{SUPA, University of Glasgow, Glasgow G12 8QQ, United Kingdom}

\author{I.~Maga\~{n}a~Hernandez}
\affiliation{University of Wisconsin-Milwaukee, Milwaukee, WI 53201, USA}

\author{H.~Qi}
\affiliation{University of Wisconsin-Milwaukee, Milwaukee, WI 53201, USA}

\author{A.~Sur}
\affiliation{Nikhef, Science Park 105, 1098 XG Amsterdam, The Netherlands}

\author{F.~Acernese}
\affiliation{Universit\`a di Salerno, Fisciano, I-84084 Salerno, Italy}
\affiliation{INFN, Sezione di Napoli, Complesso Universitario di Monte S.Angelo, I-80126 Napoli, Italy}
\author{L.~Aiello}
\affiliation{Gran Sasso Science Institute (GSSI), I-67100 L'Aquila, Italy}
\affiliation{INFN, Laboratori Nazionali del Gran Sasso, I-67100 Assergi, Italy}
\author{A.~Allocca}
\affiliation{Universit\`a di Pisa, I-56127 Pisa, Italy}
\affiliation{INFN, Sezione di Pisa, I-56127 Pisa, Italy}
\author{M.~A.~Aloy}
\affiliation{Departamento de Astronom\'{\i }a y Astrof\'{\i }sica, Universitat de Val\`encia, E-46100 Burjassot, Val\`encia, Spain}
\author{A.~Amato}
\affiliation{Laboratoire des Mat\'eriaux Avanc\'es (LMA), CNRS/IN2P3, F-69622 Villeurbanne, France}
% ...



\begin{abstract}
We perform a statistical standard siren analysis of GW170817. Our analysis does not utilize knowledge of NGC 4993 as the unique host galaxy of the optical counterpart to GW170817. Instead, we consider each galaxy within the GW170817 localization region as a potential host; combining the redshift  from each galaxy with the distance estimate from GW170817 provides an estimate of the Hubble constant, $H_0$. We then combine the $H_0$ values from all the galaxies to provide a final measurement of $H_0$. We explore the dependence of our results on the thresholds by which galaxies are included in our sample, as well as the impact of weighting the galaxies by stellar mass and star-formation rate. Considering all galaxies brighter than $0.01 L^\star_B$ as equally likely to host a BNS merger, we find $H_0= 76^{+48}_{-23}$ km s$^{-1}$ Mpc$^{-1}$ (maximum \emph{a posteriori} and 68.3\% highest density posterior interval; assuming a flat $H_0$ prior in the range $\left[ 10, 220 \right]$ km s$^{-1}$ Mpc$^{-1}$). Restricting only to galaxies brighter than $0.626 L^\star_B$ tightens the measurement to $H_0= 77^{+37}_{-18}$ km s$^{-1}$ Mpc$^{-1}$. {We show that weighting the host galaxies by stellar mass or star-formation rate provides entirely consistent results with potentially tighter constraints.} While these statistical estimates are inferior to the value from the counterpart standard siren measurement utilizing NGC 4993 as the unique host, $H_0=76^{+19}_{-13}$ km s$^{-1}$ Mpc$^{-1}$ {(determined from the same publicly available data)}, {our analysis is a proof-of-principle demonstration of the statistical approach first proposed by Bernard Schutz over 30 years ago.}

\end{abstract}


\section{Introduction} \label{sec:intro}
The first multi-messenger detection of a binary neutron star (BNS) merger, GW170817, by LIGO~\citep{2015CQGra..32g4001L} and Virgo~\citep{2015CQGra..32b4001A} enabled the first standard siren measurement of the Hubble constant, $H_0$, ushering in the era of gravitational-wave (GW) cosmology \citep{GW170817:H0,GW170817:discovery,GW170817:MMA}. This $H_0$ measurement combined the luminosity distance to the source, as measured from the GW signal \citep{Schutz}, with the known redshift of the host galaxy, NGC 4993. NGC 4993 was identified as the unique host galaxy following the discovery of an optical transient located only $\sim 10$ arcsec from NGC 4993 \citep{2017Sci...358.1556C, GW170817:DES, GW170817:MMA}. The probability of a chance coincidence between the GW signal and the optical transient was estimated to be $\lesssim 0.5\%$ \citep{GW170817:DES}, and the probability of a chance association between the optical transient and NGC 4993 is $\lesssim 0.004\%$ \citep{GW170817:H0}.

The original proposal by \cite{Schutz} to measure the Hubble constant with GW detections of compact binary mergers did not involve electromagnetic counterparts. Instead, Schutz considered bright galaxies in the GW localization region as potential hosts to the merger. Each galaxy provides a redshift that, when combined with the GW-measured luminosity distance, gives a separate estimate of $H_0$. The final $H_0$ measurement from a single event is the sum of all contributions from the individual galaxies.
The first detailed exploration of this method on simulated data, and with the first use of a galaxy catalog~\citep{2011ApJS..193...29A}, was by~\citep{DelPozzo:2012}.
An up-to-date forecast incorporating realistic detection rates, galaxy peculiar velocities, large-scale structure, and additional considerations
can be found in~\cite{Chen:H0}.
\maya{We refer to this approach of measuring $H_0$ as the ``statistical'' method~\citep{Schutz,MacLeod:2008,Petiteau:2011,Chen:H0}, compared with the ``counterpart'' method in which an electromagnetic (EM) counterpart provides a unique host galaxy association.}
In the limit where the GW event is so well-localized that there is only one potential host galaxy in the GW localization error box \citep{Chen:theone}, the statistical method reduces to the counterpart method. In the opposite limit, where the GW event is poorly localized, there are so many potential host galaxies that the distinct peaks from individual galaxies are washed out, and the $H_0$ measurement is uninformative \citep{Chen:H0}.

The statistical approach may be the only way to do standard siren science
with binary black holes, since they are not expected to have EM counterparts. We emphasize that although the statistical measurements for a given event are inferior to the counterpart case, combining many of these measurements leads to increasingly precise constraints~\citep{Schutz,DelPozzo:2012,Chen:H0, 2018arXiv180406085N}.
In ground-based gravitational wave detector networks, the rate of detection of binary black holes is significantly higher than that for neutron stars~\citep{2016PhRvX...6d1015A,GW170104,GW170817:discovery},\holz{ although the higher rate is not expected to compensate for the inferior constraints
\citep{Chen:H0}.}
Nonetheless, the black hole systems can be observed to much higher redshifts, potentially providing constraints on the evolution history of the Universe out past the turnover between dark matter and dark energy domination~\citep{DelPozzo:2012,2015ApJ...806..263D,2016Natur.534..512B,2018arXiv180510270F}. \maya{Because these systems are farther away, however, it will be a greater challenge to supply a sufficiently complete galaxy catalog.}

\daniel{In this paper we carry out a measurement of $H_0$ using the GW data from GW170817 and a catalog of potential host galaxies within the GW localization region. In other words, we explore how tight the $H_0$ measurement from GW170817 would have been if an EM counterpart had not been detected or if a unique host galaxy had not been identified. We present our methods in \S\ref{sec:methods}, a discussion of the galaxy selection in \S\ref{sec:galaxies}, a discussion of the gravitational-wave constraints in \S\ref{sec:GW}, results in \S\ref{sec:results}, and conclude in \S\ref{sec:conclusion}.
}

\section{Methods} \label{sec:methods}
We follow the statistical framework presented in \cite{Chen:H0} \citep[see also][]{DelPozzo:2012,MDC}. We include the role of GW selection effects, galaxy catalog incompleteness, galaxy luminosities, and redshift uncertainties in our analysis.
The posterior on $H_0$ given the GW and EM data, $x_{\rm GW}$ and $x_{\rm EM}$, is:
\begin{widetext}
\begin{equation}
\label{eq:posterior}
p(H_0 \mid x_{\rm GW}, x_{\rm EM}) = \frac{p_0(H_0)}{\beta(H_0)}\int p(x_{\rm GW} | \hat{D}_L(z,H_0), \Omega) p(x_{\rm EM}| z, \Omega) p_0(z, \Omega) d\Omega dz,
\end{equation}
\end{widetext}
where $\hat{D}_L(z, H_0)$ is the luminosity distance of a source at redshift $z$ for a given $H_0$~\citep[fixing other cosmological parameters to the Planck values;][]{PlanckCosmology}\footnote{For the redshifts considered here, $z \lesssim 0.05$, other cosmological parameters affect the distance-redshift relation at the sub-percent level, and so our analysis is insensitive to their precise values.}, $\Omega$ is the sky position, and $\beta(H_0)$ is a normalization term to ensure that the likelihood normalizes to 1 when integrated over all \emph{detectable} GW and EM datasets~\citep{Mandel:2016}. The term $p_0(H_0)$ represents the prior on the Hubble constant. {A detailed derivation of Equation~\ref{eq:posterior}, including the role of the normalization term $\beta(H_0)$, is provided in the Appendix.}

{As first emphasized by~\citet{Schutz}, the GW signal from a compact binary coalescence allows for a direct measurement of the distance to the source, as well as its sky location.} This measurement is represented in the GW likelihood term, $p(x_{\rm GW} | D_L, \Omega)$, which is the probability of the GW data in the presence of signal from a compact binary with parameters $D_L$ and $\Omega$ marginalized over the other parameters of the signal \maya{(including the inclination angle, component masses, spins and/or tides)}. The corresponding posterior $p(D_L, \Omega \mid x_{\rm GW}) \propto p(x_{\rm GW} | D_L, \Omega) p_0(D_L, \Omega)$ is summarized in the  3-dimensional sky map, which provides a fit to the posterior samples provided by the GW parameter estimation pipeline LALInference \citep{LALInference, GtD, GtDSupplement}. For this analysis, we use the publicly released 3-dimensional sky map from~\cite{SourceProperties} (see~\S \ref{sec:GW} and~\S\ref{sec:results}). To get the likelihood from the posterior probability, we must first divide out the default ``volumetric" distance prior, $p_0(D_L, \Omega) \propto D_L^2$.

Meanwhile, the EM likelihood term $p(x_{\rm EM} \mid z, \Omega)$ is the probability of the electromagnetic data in the presence of signal from a compact binary with parameters $z$ and $\Omega$. In the absence of an EM counterpart and/or a host galaxy identification, we assume the measurement $p(x_{\rm EM} | z, \Omega)$ is completely uninformative, and set:
\begin{equation}
p(x_{\rm EM} | z, \Omega) \propto 1.
\end{equation}
In this case, the redshift information enters only through the prior term, $p_0(z, \Omega)$, which we take to be a galaxy catalog. The detection of an electromagnetic counterpart typically results in $p(x_{\rm EM} \mid z, \Omega)$ being strongly peaked around some $\hat{\Omega}$ allowing the identification of a host galaxy. We note that in some cases an optical transient may be identified, but it may not be possible to uniquely identify the associated host galaxy. In these circumstances one could perform a pencil-beam survey of the region surrounding the transient (e.g., at distances of $\lesssim100\,$kpc from the line-of-sight to the transient), and sharply reduce the relevant localization volume~\citep{Chen:H0}. This reduces the number of potential host galaxies, and thereby improves the measurement.

The galaxy catalog term $p_0(z,\Omega)$ is given by:
\begin{equation}
\label{eq:zprior_incomplete}
p_0(z, \Omega) = f p_{\rm cat}(z, \Omega) + (1-f)p_{\rm miss}(z, \Omega),
\end{equation}
where $p_{\rm cat}$ is a catalog of known galaxies, $p_{\rm miss}$ represents the distribution of missing galaxies, and $f$ denotes the overall completeness fraction of the catalog. The contribution from the known galaxies is:
\begin{equation}
\label{eq:zcomplete}
p_{\rm cat}(z, \Omega) = \sum_i^{N_\mathrm{gal }}w_i N(\bar{z}_i,\sigma_z; z)N(\bar{\Omega}_i, \sigma_\Omega; \Omega),
\end{equation}
where $\bar{z}_i, \bar{\Omega}_i$ denotes the (peculiar velocity-corrected) ``Hubble'' redshifts and sky coordinates of all galaxies in the catalog, and $N(\mu, \sigma; x)$ denotes the normal probability density function with mean $\mu$ and standard deviation $\sigma$ evaluated at $x$. \maya{To account for peculiar velocity uncertainties, which can be significant for nearby sources, we assume that the true Hubble velocity is normally distributed about the measured Hubble velocity with an uncertainty of $c \sigma_z$~\citep{Scolnic:2017b}.} On the other hand, the uncertainty on the sky coordinates of galaxies in the catalog is negligible for our purposes, so we always approximate $N(\bar{\Omega}, \sigma_\Omega;\Omega)$ by $\delta(\bar{\Omega}-\Omega)$.

The weights $w_i$ can be chosen to reflect the {\em a priori}\/ belief that a galaxy could host a gravitational-wave source. For example, setting all weights to $w_i = \frac{1}{N_\mathrm{gal}}$ corresponds to equal probability for each galaxy to host a gravitational wave source. In general, since we might expect that the BNS rate is traced by some combination of stellar mass and/or star formation rate~\citep{phinney1991rate, Leibler:2010,Fong:2013,2018MNRAS.474.2937C}, we may assign unequal weights to galaxies based on these (or any other relevant observable) quantities, ensuring that the weights sum to unity. In the following, we use a galaxy's B-band luminosity as a proxy for its star formation rate, and its K-band luminosity as a proxy for its total stellar mass \citep{Bell:2003,GtD}; these are very rough estimates, but serve to demonstrate the method. \maya{In these cases, we apply weights proportional to B-band or K-band luminosity, $w_i \propto L_B^i$ or $w_i \propto L_K^i$}, and explore the dependence of the result on these of weightings.
\dan{We should probably have a discussion somewhere about bias due to GRBs/KN somehow preferentially selecting face-on galaxies?}

\maya{To calculate the term $p_{\rm miss}$ in Equation~\ref{eq:zprior_incomplete}, we assume that on large scales, the distribution of galaxies, $p_0(z, \Omega)$ is uniform in comoving volume.} Let $p_{\rm vol}(z, \Omega)$ denote the cosmologically homogeneous and isotropic distribution normalized over the volume contained in the range $z_\mathrm{min} < z < z_\mathrm{max}$ considered in our analysis. (The result does not depend on the choice of $z_\mathrm{min}$ or $z_\mathrm{max}$ provided that the interval encompasses all possible redshifts of the source for all allowed values of $H_0$.) Assuming all galaxies are weighted equally, the distribution of missing galaxies is written as:
\begin{equation}
\label{eq:pmissing}
p_{\rm miss} (z, \Omega) = \frac{\left[ 1-P_{\rm complete}(z) \right] p_{\rm vol}(z,\Omega)}{(1-f)},
\end{equation}
where $P_{\rm complete}(z)$ is the probability that a galaxy at redshift $z$ is in the catalog, and the completeness fraction $f$ is given by:
\begin{equation}
f = \int_{z_\mathrm {min}}^{z_\mathrm{max}} P_{\rm complete}(z) p_{\rm vol}(z, \Omega) \, dz \, d\Omega.
\end{equation}

We can similarly add galaxy weightings to an incomplete catalog by computing the luminosity distribution of the ``missing galaxies'' as a function of redshift, $p(L \mid z, \rm missing)$. We calculate this distribution by assuming that the luminosities of the missing galaxies together with those in the catalog are distributed according to the Schechter function. Then, the weights of the missing galaxies are given by:
\begin{equation}
w(z) \propto \int L p(L \mid z, \mathrm{missing}) \, dL,
\end{equation}
and, weighting each missing galaxy by its luminosity, Equation~\ref{eq:pmissing} becomes:
\begin{equation}
p_{\rm miss} (z, \Omega) \propto
w(z) p_\mathrm{miss}(z, \Omega).
\end{equation}
Note that we have assumed that the coverage of the catalog is uniform over the sky and so $P_{\rm complete}$ is independent of $\Omega$. (This is true over the relevant sky area for the present analysis, but the method can be easily generalized to add an $\Omega$-dependence.)
Alternate approaches of taking into account the incompleteness of galaxy catalogs are being explored in \cite{MDC}. However, in the present case of a single nearby source where the catalog is largely complete, the differences in results from the various approaches are small, and in particular, well within the statistical uncertainties.
In section~\ref{sec:galaxies}, the completeness function $P_\mathrm{complete}$ is estimated for the galaxy catalog used in the analysis.

\begin{figure}
\includegraphics[width=0.5\textwidth]{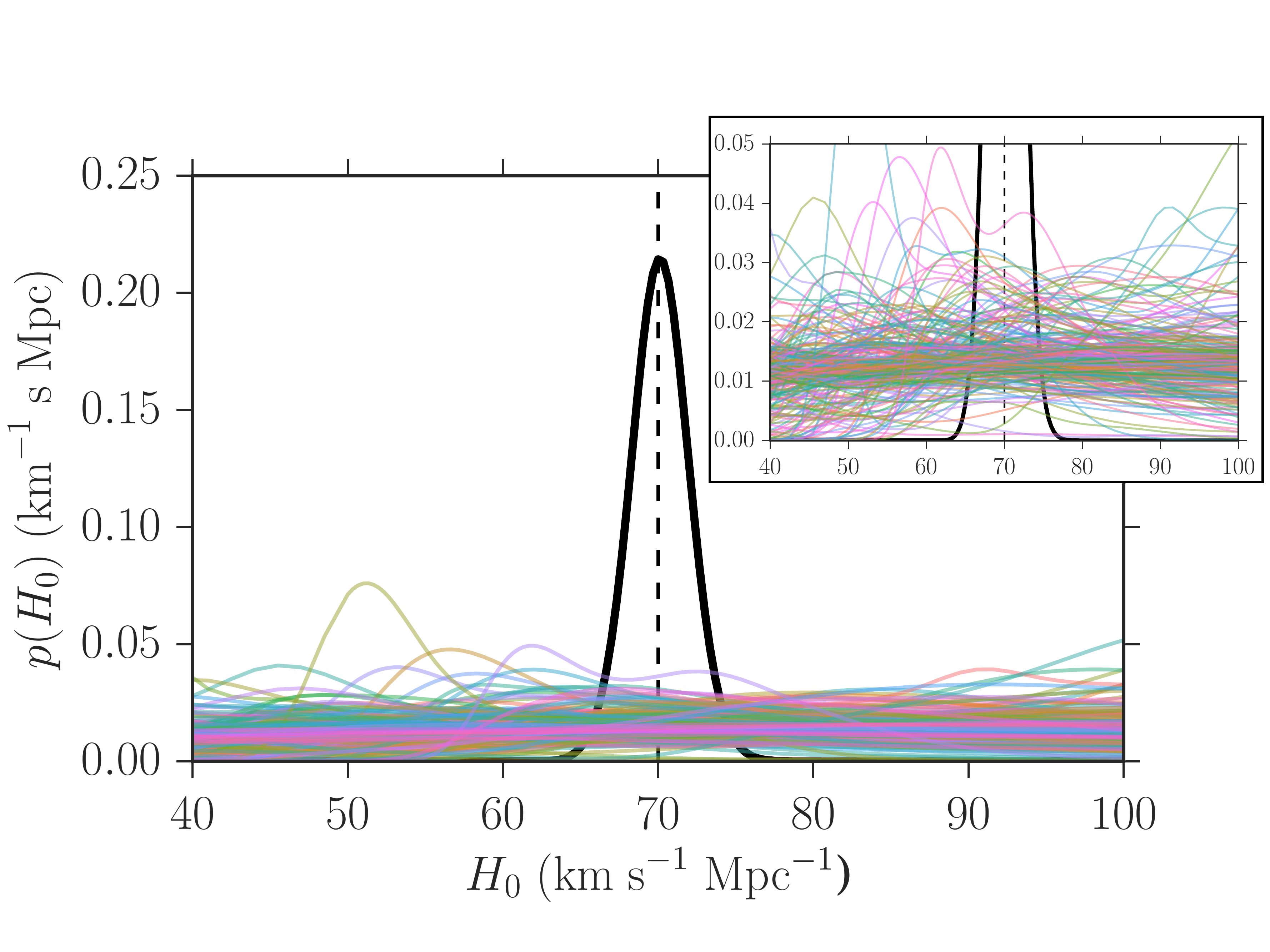}
\caption{\label{fig:pH0MICE} Projected $H_0$ constraints using the statistical method on a sample of 249 simulated BNS detections and the MICE mock galaxy catalog. The thin colored lines show the $H_0$ posteriors from individual events, while the solid black curve shows the combined posterior. The prior is assumed to be flat in all cases. The dashed black line shows the injected value, $H_0 = 70$ km s$^{-1}$ Mpc$^{-1}$.}
\end{figure}

\rr{To demonstrate the statistical method, we apply the analysis described above to 249 simulated BNS GW detections from the First Two Years (F2Y) catalog~\cite{F2Y} and the MICE simulated galaxy catalog~\citep{MICEI,MICEII,MICEIII,MICEIV,MICEV,Carretero:2017zkw}. We assign each  BNS detection from the F2Y 2016 scenario (roughly corresponding to O2) to a galaxy in the MICE catalog with a redshift that matches the injected distance and assumed $H_0$ value ($H_0 = 70$ km s$^{-1}$ Mpc$^{-1}$). For each event, we rotate the sky coordinates of the galaxies in the catalog so that the sky position of the host galaxy matches the true sky position of the BNS injection. We then carry out the statistical method using the 3-dimensional sky map for each mock BNS and the galaxies in MICE, assuming no peculiar velocities or incompleteness, and assigning weights to the galaxies in MICE so that the redshift distribution matches the injected redshift distribution of the F2Y dataset, $p(z) \propto z^2$. This last step is necessary in order to ensure that the selection effects are incorporated consistently between the injections and the likelihood. The results are shown in Fig.~\ref{fig:pH0MICE}. Even in the best-case scenario of perfectly-known galaxy redshifts and a complete catalog, the $H_0$ posteriors from most individual events are nearly flat over the prior range. Combining the 249 individual events, the final $H_0$ posterior is $H_0 = 70.1^{+1.9}_{-1.9}$ km s$^{-1}$ Mpc$^{-1}$ (68.3\% credible interval), corresponding to a convergence rate of $\sim40\%/\sqrt{N}$, consistent with~\cite{Chen:H0}. \daniel{As is visible in the Figure, we confirm that the method is unbiased, with the result for large numbers of detections approaching the true value of $H_0$.} We note that most of the simulated detections in the F2Y dataset have much larger localization volumes than GW170817, which was an unusually loud, nearby source that was detected while all three detectors were operational. Therefore, we expect the statistical $H_0$ measurement from GW170817 to be unusually informative compared to an average event. We quantify this expectation in Section~\ref{sec:results}.}

\section{Galaxy catalogs}\label{sec:galaxies}
\begin{figure}
\includegraphics[width=0.5\textwidth]{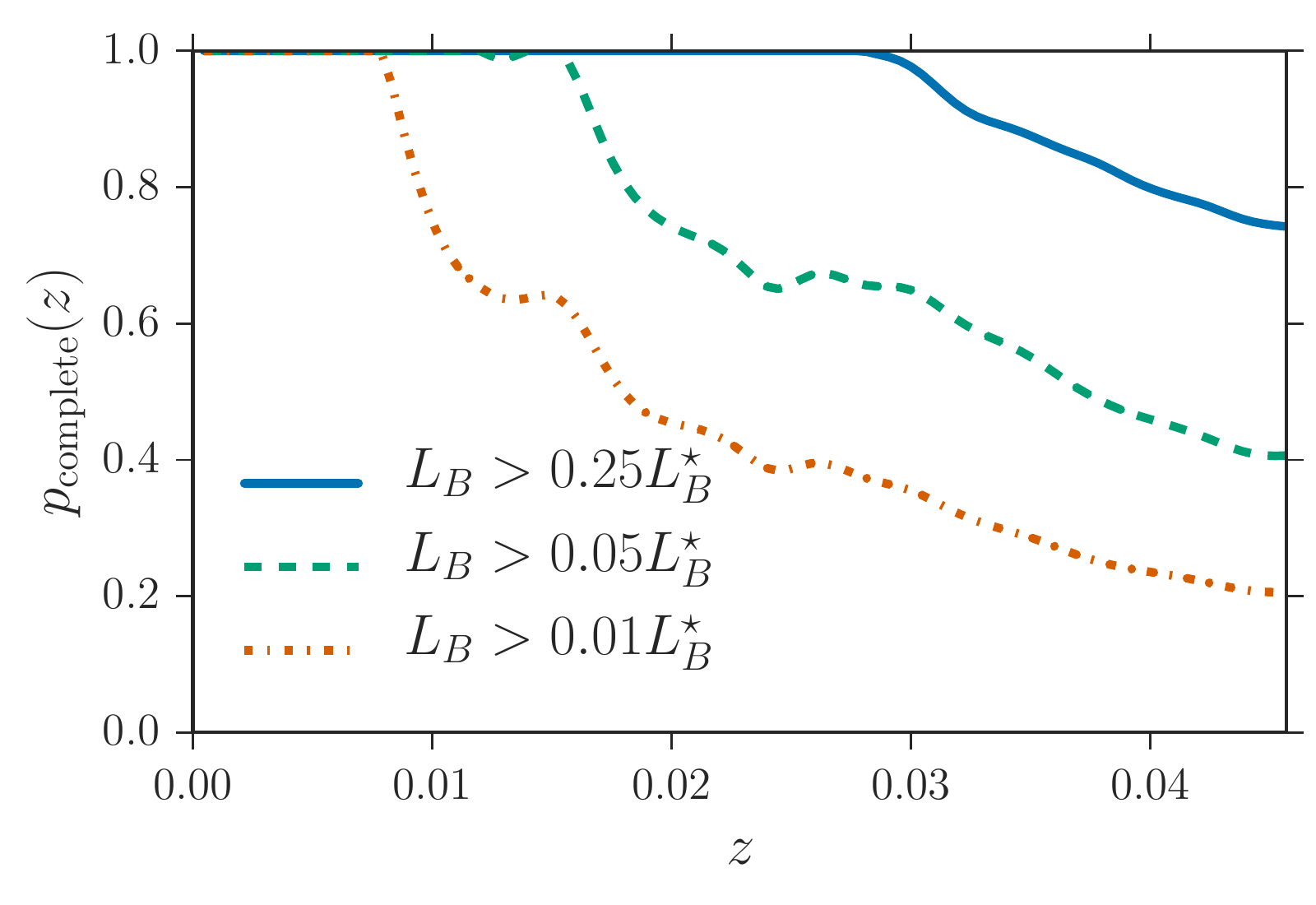}
\caption{\label{fig:pzcomplete} Completeness of the GLADE catalog as a function of redshift for galaxies brighter than $0.25 L^\star_B$ (solid blue curve), $0.05 L^\star_B$ (dashed green curve), and $0.01 L^\star_B$ (dot-dahsed orange curve), calculated by comparing the redshift distribution of galaxies in GLADE to a distribution that is constant in comoving volume. For galaxies brighter than $0.626 L^\star_B$, GLADE is complete across the entire redshift range shown.}
\end{figure}
To measure $H_0$ statistically with GW170817, we use version 2.3 of the GLADE galaxy catalog to construct our redshift prior in Equation~\ref{eq:zprior_incomplete} \citep{GLADE}. GLADE provides galaxy redshifts in the heliocentric frame, corrected for peculiar motions using the peculiar velocity catalog of \cite{Carrick:2015}. For galaxies which are also listed in the group catalog of \cite{KTGroups}, as identified by a common Principal Galaxy Catalog (PGC) identifier, we apply an additional correction to correct their velocities to the radial velocity of the group. We assume the group velocity is given by the unweighted mean of the velocities of all member galaxies, although we note that for the dominant group containing NGC 4993, careful group modeling has been done \citep{Hjorth:2017}. Finally, we correct all heliocentric velocities to the reference frame of the cosmic microwave background~\citep{Hinshaw:2009} and assign a 200 km/s Gaussian uncertainty to the ``Hubble velocity'' of each galaxy in the catalog \citep[corrected by all peculiar motions;][]{Carrick:2015,Scolnic:2017b}.

GLADE also provides luminosity information for galaxies, listing apparent magnitudes in the B-, J-, H-, and K-bands. We use the reported B-band luminosities to characterize the completeness of the catalog (a small fraction of galaxies do not have B-band apparent magnitudes reported in the catalog; we remove these galaxies from our analysis, assuming that their magnitudes are below our adopted luminosity cutoff). Following \cite{Gehrels:2016} and \cite{Arcavi:2017}, we adopt B-band Schechter function parameters $\phi^\star = 5.5 \times 10^{-3} h_{0.7}^3$ Mpc$^{−3}$, $\alpha_B = -1.07$, $L^\star_B = 2.45 \times 10^{10} h_{0.7}^{−2} L_{B,\odot}$ throughout. The corresponding characteristic absolute magnitude is $M^\star_B = -20.47 + 5 \log_{10}h_{0.7}$. We will also consider the K-band magnitudes reported in GLADE when applying galaxy weights, and we use the K-band Schechter function parameters of $\alpha_K = -1.02$, $M^\star_K = -23.55 + 5 \log_{10}h_{0.7}$ \citep{Lu:2016}.

\begin{figure}[b]
\includegraphics[width=0.5\textwidth]{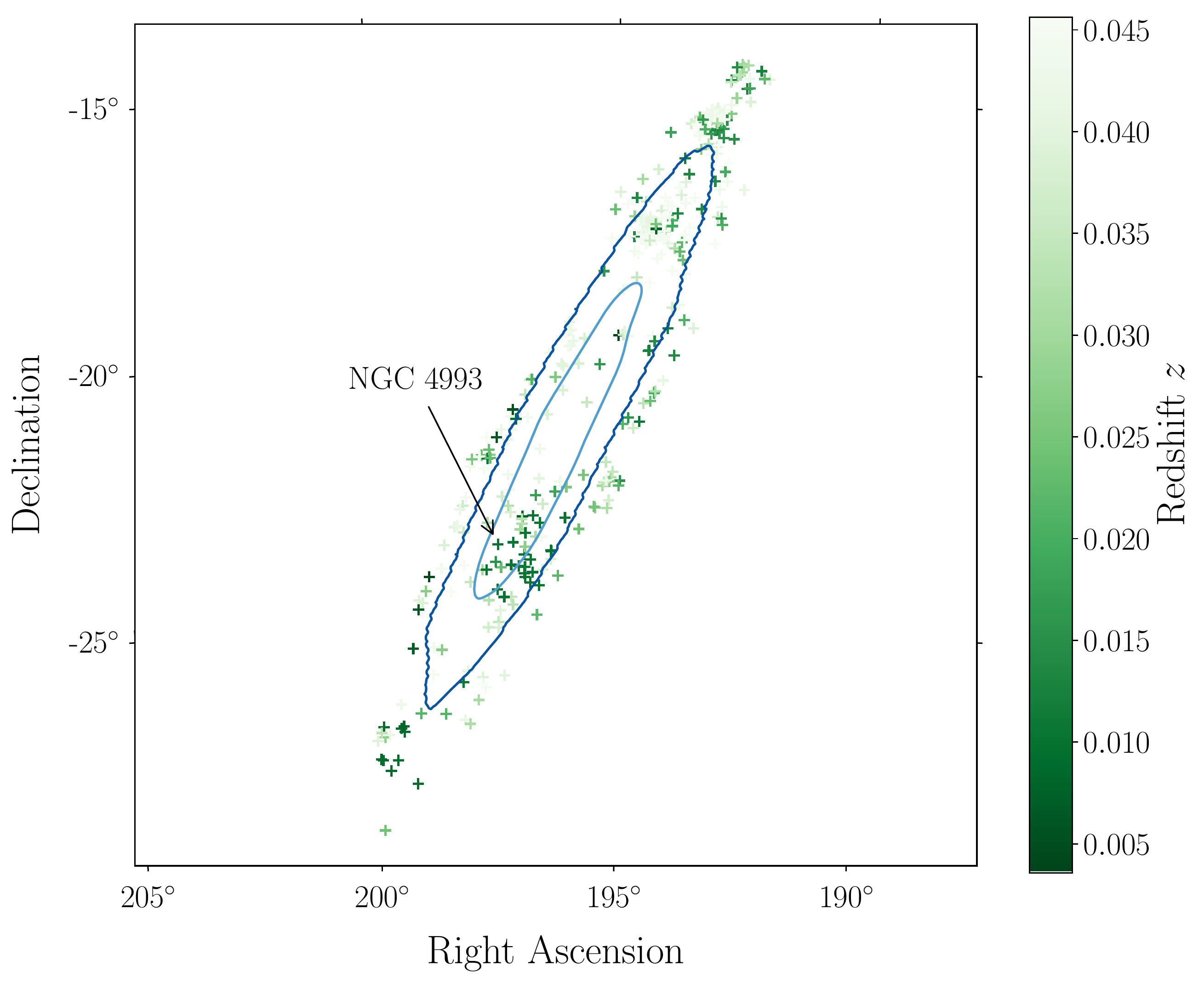}
\caption{\label{fig:skymap} Two-dimensional localization region of GW170817 (blue contours) with the sky coordinates of the 408 GLADE galaxies (green crosses) within the 99\% localization area and the redshift range $0 < z \lesssim 0.046$ (for an $H_0$ prior range of $H_0 \in \left[ 10, 220 \right]$ km s$^{-1}$ Mpc$^{-1}$). The light and dark blue contours enclose the 50\% and 90\% probability regions, respectively, and the shading of the galaxy markers denotes their redshifts, corrected for peculiar and virial motions as described in the text.}
\end{figure}

Figure~\ref{fig:pzcomplete} summarizes the completeness of GLADE as a function of redshift. We find that GLADE is complete up to redshifts $z \sim 0.06$ for galaxies brighter than $\sim 0.626 h_{0.7}^{-2} \ L^\star_B$, corresponding to about 0.66 of the Milky Way luminosity for $h_{0.7} = 1$. Galaxies brighter than $\sim 0.626 L^\star_B$ make up half of the total luminosity for the given Schechter function parameters.
We find that for $z \lesssim 0.03$, GLADE is complete for galaxies down to 2.5 times dimmer, or $\sim 0.25 L^\star_B$, corresponding to $M_B = -18.96 + 5\log_{10}h_{0.7}$ \citep[see also Figure~2 of ][]{Arcavi:2017}. Such galaxies make up 75\% of the total B-band luminosity. If we consider galaxies down to $\sim 0.05 L^\star_B$ ($M_B = -17.22 + 5\log_{10}h_{0.7}$), GLADE is $\sim 70\%$ complete at $z \sim 0.03$, and even if we consider galaxies down to $\sim 0.01 L^\star_B$ ($M_B = -15.47 + 5\log_{10}h_{0.7}$), including 99\% of the total B-band luminosity, GLADE is $\gtrsim 80\%$ complete for $z \lesssim 0.01$, and $\sim 40\%$ complete at $z \sim 0.03$. In the K-band, we find that with our assumed K-band Schechter function parameters, GLADE is complete up to $z \sim 0.045$ for galaxies with $L_K > 0.36 L^\star_K$, which contain 70\% of the total K-band luminosity, and up to $z \sim 0.03$ for galaxies with $L_K > 0.1 L^\star_K$, which contain 90\% of the total luminosity. For galaxies brighter than $L_K = 0.005 L^\star_K$, which make up more than 99\% of the total K-band luminosity, GLADE is $\sim 70\%$ complete at $z = 0.01$.

\section{Source Localization and Distance}\label{sec:GW}

\begin{figure}
\includegraphics[width=0.5\textwidth]{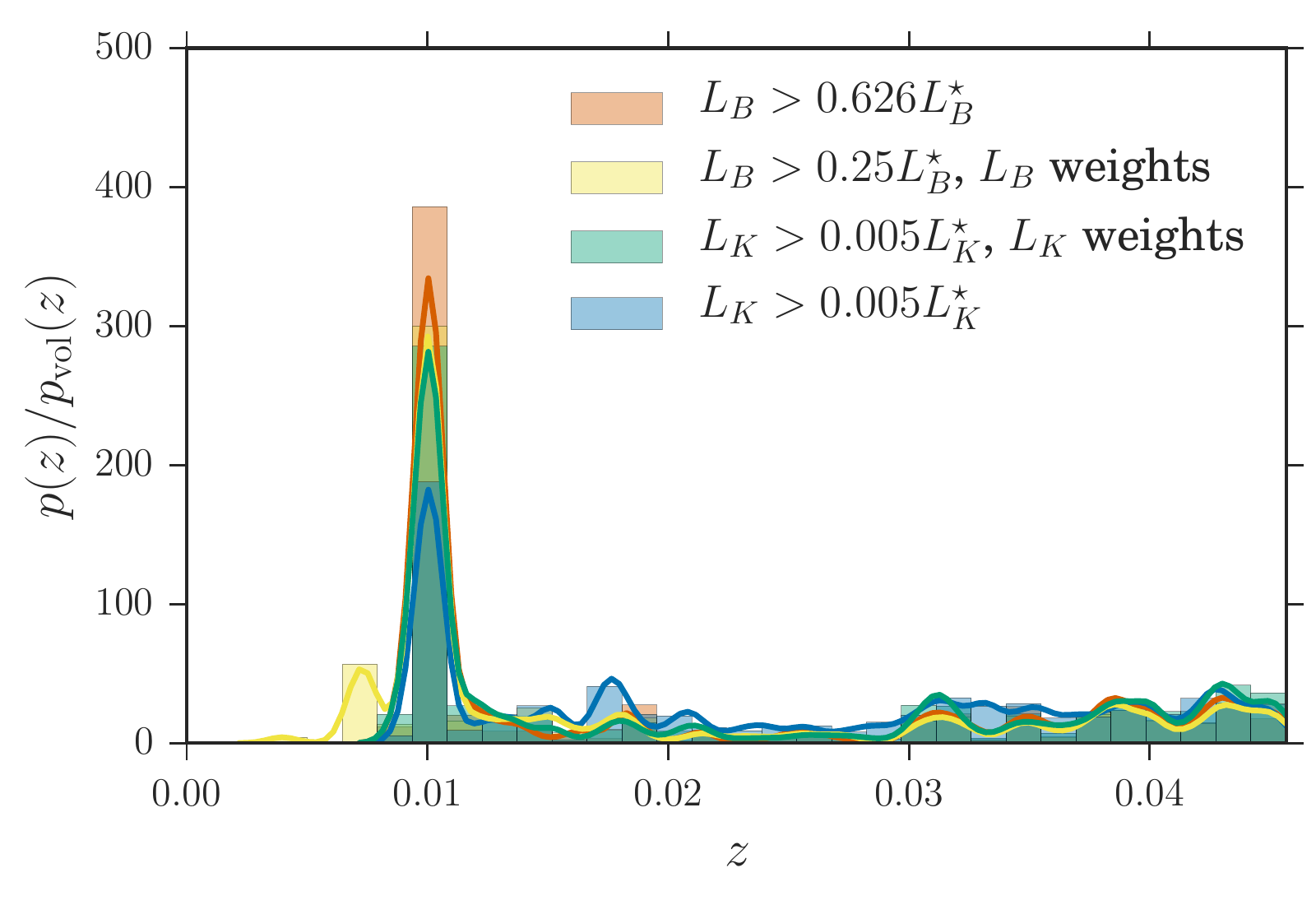}
\caption{\label{fig:galaxyzs} Probability distribution of the redshifts of potential hosts to GW170817 weighted by the GW sky map probability, $p(z) = \int p(x_\mathrm{GW} \mid \Omega)p_0(z, \Omega) d\Omega$, compared to a uniform in comoving volume distribution of galaxies, $p_{\rm vol}(z)$. For the orange histogram, we include all galaxies in the catalog brighter than $0.626 L^\star_B$.  For galaxies brighter than $0.626 L^\star_B$, the catalog is complete over the redshift range. However, when we lower the luminosity cutoff to $0.25 L^\star_B$ (yellow histogram) or $0.005 L^\star_K$ (green and blue), we must account for catalog incompleteness at higher redshifts \maya{by considering the redshift and luminosity distributions of the missing galaxies} (see \S\ref{sec:methods}). The yellow (green) histogram additionally weights each galaxy by its B-band (K-band) luminosity. If the ratio $p(z) / \ {p_{\rm vol}}\left(z\right)$ were completely flat, we would expect an uninformative $H_0$ measurement in which our posterior recovers our prior. However, in all instances there is a dominant peak at $z \sim 0.01$, suggesting that the resulting $H_0$ measurement will be informative. Adding in luminosity weights, especially in the K-band, makes the peak more dominant.}
\end{figure}

From the GW data alone, GW170817 is the best-localized GW event to date. The original analysis by the LIGO-Virgo collaboration \citep{GW170817:discovery} reported a 90\% localization area of 28 deg$^2$ and a 90\% localization volume of 380 Mpc$^3$~\citep[assuming Planck cosmology;][]{PlanckCosmology}, while the most recent analysis \citep{SourceProperties} improves the 90\% localization area to 16 deg$^2$ and the 90\% volume to 215 Mpc$^3$. {We use this updated 3-dimensional sky map \citep{GtD, GtDSupplement} from~\cite{SourceProperties} throughout.\footnote{With the data release accompanying~\cite{SourceProperties}, the LIGO-Virgo collaboration has made the {3-dimensional data behind this} sky map publicly available at the following url: \url{https://dcc.ligo.org/DocDB/0150/P1800061/009/figure_3.tar.gz}}} Figure~\ref{fig:skymap} shows the 2-dimensional sky map together with the galaxies in the GLADE catalog within the localization region. Figure~\ref{fig:galaxyzs} shows that, although there are a total of $408$ galaxies within the 99\% localization area (see Figure~\ref{fig:skymap}), most of the galaxies with high sky-map probability come in a few distinct groups: \maya{a dominant group at $z \sim 0.01$ regardless of the assumed luminosity threshold, followed by a secondary group at $z \sim 0.006$ containing only moderately faint galaxies}. Therefore, there are only a few distinct redshifts that can possibly correspond to the measured distance of GW170817, and we expect that combining the galaxy catalog with the GW localization will yield an informative measurement of the Hubble constant.

\begin{figure}[b]
\includegraphics[width=0.5\textwidth]{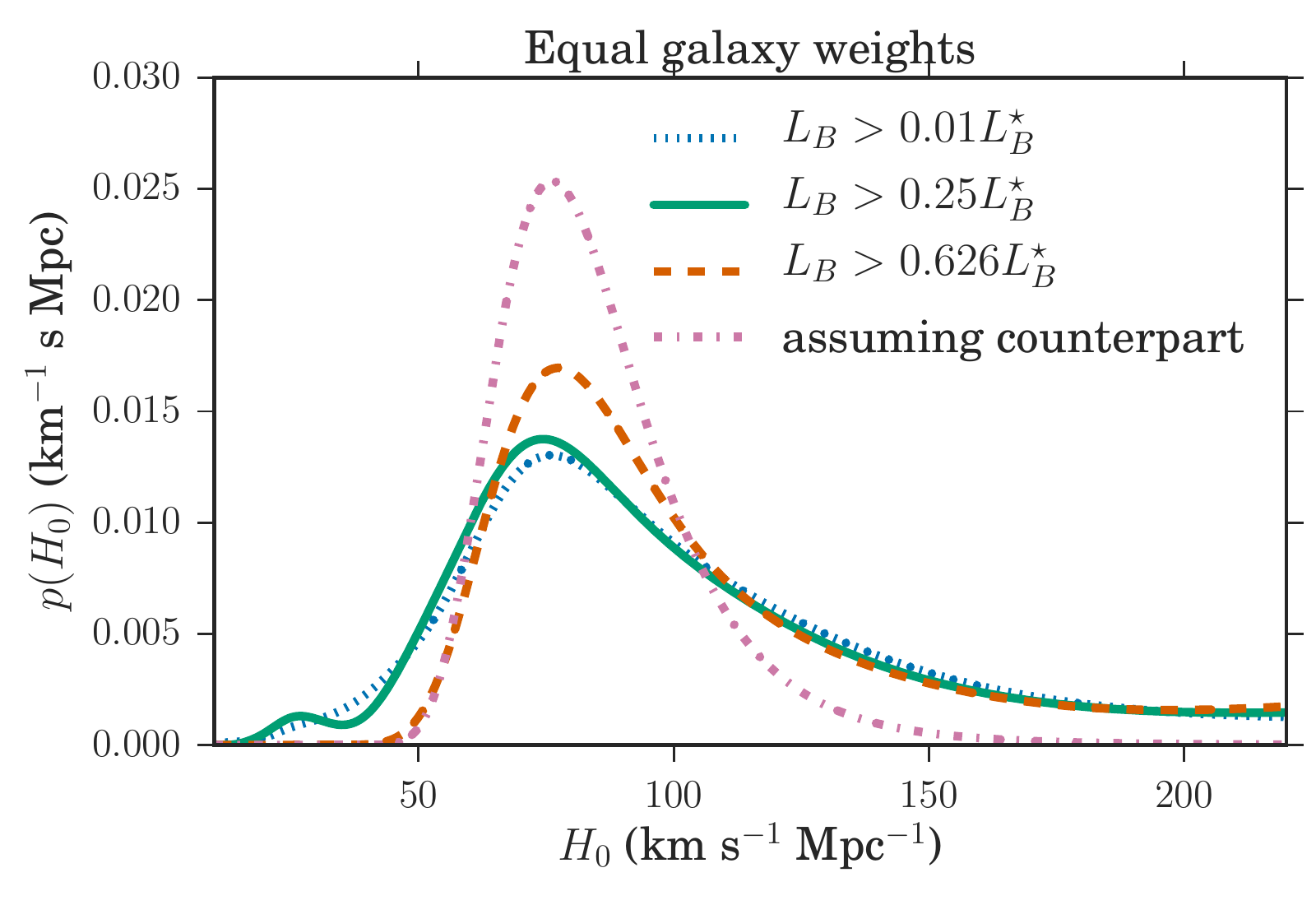}
\caption{\label{fig:H0post}
Posterior probability of $H_0$ under various assumptions regarding the potential host galaxy. We adopt a flat $H_0$ prior in the range $H_0 \in \left[ 10, 220 \right]$ km s$^{-1}$ Mpc$^{-1}$. For the dashed orange curve, we assume that only galaxies brighter than $0.626 L^\star_B$ (containing 50\% of the total luminosity) can host BNS events, meaning that the galaxy catalog is complete over the relevant redshift range. The solid green curve lowers the luminosity cutoff to $0.25 L^\star_B$ (containing 75\% of the total luminosity), and accounts for the mild incompleteness of the catalog above redshifts $z \sim 0.03$. The dotted blue curves incorporate all galaxies brighter than $0.01 L^\star$ (containing 99\% of the total luminosity), accounting for the incompleteness of faint galaxies at redshifts $z \gtrsim 0.01$. The dot-dashed pink curve shows the $H_0$ measurement assuming the host galaxy is known to be NGC 4993.}
\end{figure}

\daniel{The 3-dimensional sky map also provides an approximation to the luminosity distance posterior along each line-of-sight. As usual, the distance to GW170817 is determined directly from the gravitational waves, and is {calibrated by general relativity~\citep{Schutz}}. No distance ladder is required.}

\section{Results}\label{sec:results}

\begin{figure*}
\includegraphics[width=\textwidth]{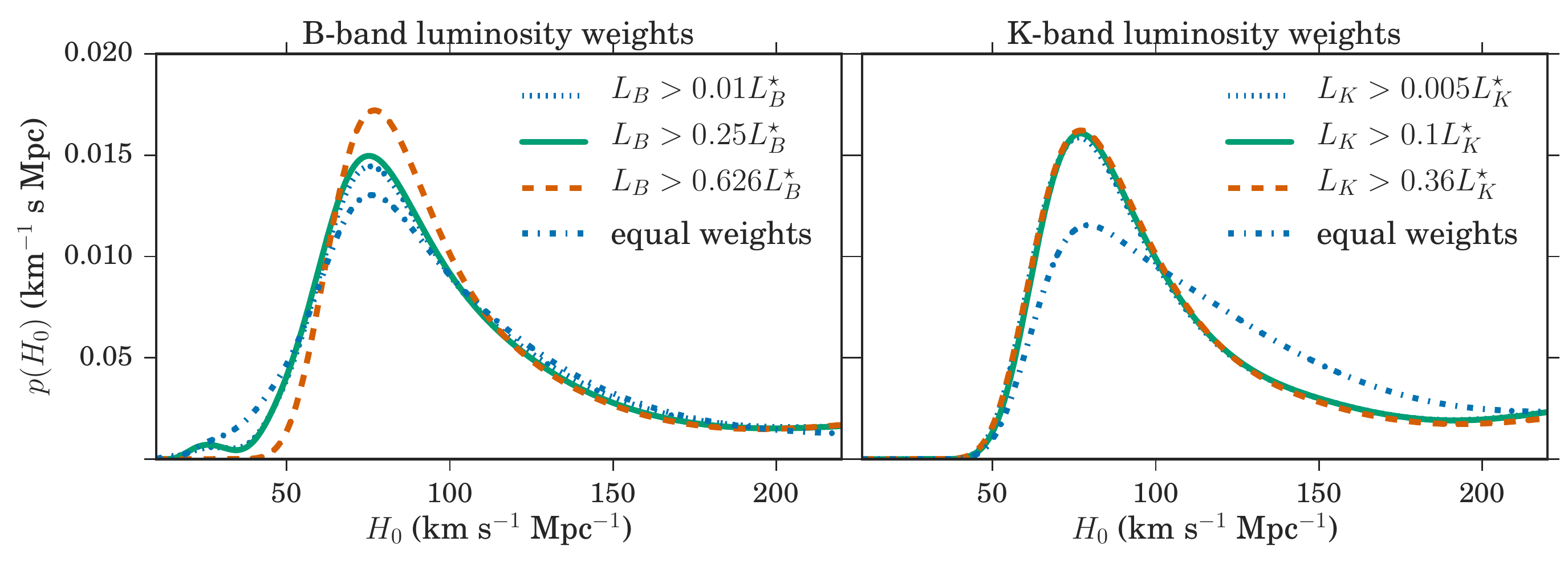}
\caption{\label{fig:H0post_lumweights}
Posterior probability of $H_0$, weighting all galaxies in the volume by their B-band luminosities, corresponding roughly to weighting by star-formation rate (left),  or K-band luminosities, corresponding roughly to weighting by stellar mass (right). We have applied the necessary completeness correction (see~\S\ref{sec:methods}). The blue dashed-dot curve shows all galaxies brighter than $0.01 L^\star_B$ in B-band (left) or $0.005 L^\star_K$ in K-band (right) with equal weights for comparison. Weighting galaxies by their K-band luminosities brings all the curves into very close agreement, because many galaxies in the group at $z \sim 0.01$ have brighter than average K-band luminosities (brighter than $1.5 L^\star_K$) and thus dominate the K-band weighted population and contain the majority of the stellar mass.}
\end{figure*}

\daniel{We combine the GW distance posterior for GW170817 with the redshift for each potential host galaxy within the localization region. As detailed in \S\ref{sec:methods}, each galaxy produces a posterior probability for $H_0$, and we combine these estimates among all the galaxies in the localization region to arrive at a final estimate for $H_0$. We adopt a flat prior in $H_0$ over the range 10--220$\,$km s$^{-1}$ Mpc$^{-1}$. The results are presented in Figure~\ref{fig:H0post}. Because the galaxies are predominantly found in one galaxy group at $z\sim0.01$, the $H_0$ posterior shows a clear peak at $H_0 \approx 76\,$km s$^{-1}$ Mpc$^{-1}$. And because NGC 4993, the true galaxy host of GW170817, is a member of the group at $z\sim0.01$, we should not be surprised to learn that the peak in $H_0$ is consistent with the $H_0$ estimate from the GW170817 standard siren measurement including the counterpart~\citep{GW170817:H0}.
\maya{Because this analysis has been performed on a 3-dimensional sky map using an approximation to the distance posteriors, rather than using the full 3-dimensional LIGO/Virgo posteriors, the results do not agree precisely with those of~\citet{GW170817:H0}, and in particular, the position of the peak in Figure~\ref{fig:H0post} is at $H_0=76\,$km s$^{-1}$ Mpc$^{-1}$ instead of $H_0=70\,$km s$^{-1}$ Mpc$^{-1}$. This is because our 3-dimensional sky map approximates the distance posterior along each line-of-sight by a simple 2-parameter Gaussian fit \citep[see Eq.~1 of][]{GtD}, which is an imperfect approximation to the true, asymmetric distance posterior~\citep{2017ApJ...840...88C,DelPozzo:2018}. On the other hand, the analysis in~\cite{GW170817:H0} utilizes the full distance posterior along the line-of-sight to NGC 4993 rather than the Gaussian approximation.}
}

\daniel{Figure~\ref{fig:H0post} shows four different posterior probability distributions, each using a different threshold for the galaxy catalog. In the ``assuming counterpart'' case, NGC 4993 (which is assumed to be the true host galaxy to GW170817) is given a weight of 1, and all the other galaxies in the localization volume are given a weight of 0. We find $H_0 = 76^{+19}_{-13}\,$km s$^{-1}$ Mpc$^{-1}$ (maximum \emph{a posteriori} and 68.3\% highest density posterior interval) for our default flat prior, or $H_0 = 74^{+18}_{-12}\,$km s$^{-1}$ Mpc$^{-1}$ for a flat-in-log prior (the prior choice in \cite{GW170817:H0}).
This peak is slightly shifted compared to the result presented in \cite{GW170817:H0}, $H_0 = 70^{+12}_{-8}\,$km s$^{-1}$ Mpc$^{-1}$, due to the usage of the Gaussian fit to the distance posterior found in the 3-dimensional sky map as discussed above.
}

\daniel{The other curves in Figure~\ref{fig:H0post} assume different limiting thresholds for what constitutes a potential host galaxy. For a luminosity threshold of $L > 0.626 L_B^\star$, we find $H_0 = 77^{+37}_{-18}$ km s$^{-1}$ Mpc$^{-1}$~\footnote{The upper limits of the 68.3\% highest density posterior intervals that we report here are especially sensitive to the upper limit we consider for the $H_0$ prior, 220 km s$^{-1}$ Mpc$^{-1}$}. As the threshold is lowered, additional galaxies fall into the sample, and the $H_0$ posterior is broadened. For a limiting B-band magnitude of $0.25L_B^\star$, we need to account for the incompleteness of the galaxy catalog at redshifts $z\gtrsim0.03$, and for $0.01L_B^*$, we need to account for the incompleteness at $z\gtrsim0.01$, as described in \S\ref{sec:methods}. The incompleteness correction leads to a slight additional broadening of the $H_0$ posterior, but the clear peak at $H_0\approx 76\,$km s$^{-1}$ Mpc$^{-1}$ remains: we find $H_0 = 74^{+45}_{-24}$ km s$^{-1}$ Mpc$^{-1}$ for a luminosity threshold of $L > 0.25 L_B^\star$ $H_0 = 76^{+48}_{-23}$ km s$^{-1}$ Mpc$^{-1}$ for a luminosity threshold of $L > 0.01 L_B^\star$. This peak is the result of the galaxy group at $z\sim0.01$, of which NGC 4993 is a member.}

The curves in Figure~\ref{fig:H0post_lumweights} weight each galaxy by its B-band luminosity (a proxy for its recent star formation history; right) or its K-band luminosity (a proxy for its stellar mass; left). The peak at $H_0 \approx 76$ km s$^{-1}$ Mpc$^{-1}$ becomes more pronounced when galaxies are weighted by their luminosity, as the group containing NGC 4993 consists of many bright, mostly red galaxies. If we assume that the probability of hosting a BNS merger is proportional to a galaxy's B-band luminosity, the posterior on $H_0$ tightens from
$H_0\in[53,124]$ km s$^{-1}$ Mpc$^{-1}$ (68.3\% highest density posterior interval) when applying equal weights to all galaxies brighter than 0.01 $L_B^\star$ to $H_0\in[54,120]$ km s$^{-1}$ Mpc$^{-1}$. Applying K-band luminosity weights to galaxies brighter than 0.005 $L_K^\star$, the 68.3\% posterior interval tightens from $H_0\in[61,137]$ km s$^{-1}$ Mpc$^{-1}$ to $H_0\in[57,118]$ km s$^{-1}$ Mpc$^{-1}$.
{Although these results are suggestive that weighting by stellar-mass or star-formation rate may lead to faster convergence, the properties of BNS host galaxies are still uncertain, and it is impossible to establish this definitively with a single event. As the source sample increases it is expected to relate to some combination of these quantities, and incorporating these trends will lead to improvements in the statistical $H_0$ analysis.}

\rr{In order to quantify the degree of information in the GW170817 $H_0$ posterior compared to an ``average'' event as expected from the F2Y dataset, we consider the difference in the Shannon entropy between the flat prior and the posterior~\citep[see Appendix;][]{Shannon}. We compare this measure of information for the statistical GW170817 $H_0$ posterior to the individual statistical $H_0$ posteriors from each of the simulated BNS events in Section~\ref{sec:methods}. We find that for a flat prior in the (relatively narrow) range $H_0\in[40,100]$ km s$^{-1}$ Mpc$^{-1}$, the information gained by applying the statistical method to GW170817 is 0.34 bits.
Meanwhile, the median information in an individual posterior shown in Figure~\ref{fig:pH0MICE} is only 0.047 bits, so that GW170817 is in the top $\sim90\%$ of informative events, even under optimistic assumptions for the simulated detections (i.e. complete galaxy catalogs and perfect redshift measurements). \daniel{As expected, GW170817 provides an unusually good statistical $H_0$ constraint.}

For the purposes of this calculation, we use the K-band luminosity-weighted $L_K > 0.1 L_K^\star$ posterior shown in the right panel of Figure~\ref{fig:H0post_lumweights} as a representative posterior for the statistical GW170817 $H_0$ measurement. Over the wider prior $H_0\in[10,220]$ km s$^{-1}$ Mpc$^{-1}$ shown, the information difference between the posterior and the prior is 0.67 bits. The counterpart GW170817 $H_0$ measurement (dot-dashed pink curve in Figure~\ref{fig:H0post}) has an information gain of 1.55 bits with respect to the wide prior.}

\section{Conclusion}\label{sec:conclusion}

We perform a statistical standard siren measurement of the Hubble constant with GW170817. This analysis is the first application of the measurement originally proposed over 30 years ago by \citet{Schutz}.
We find that the excellent localization of GW170817, together with the large scale structure causing galaxies to cluster into distinct groups, enables an informative measurement of $H_0$ even in the absence of a unique host galaxy identification.
Including generic and flexible assumptions regarding the luminosities of BNS host galaxies, we find a peak at $H_0 \approx 76$ km s$^{-1}$ Mpc$^{-1}$ at $\sim 2.4$--$3.7$ times the prior probability density.
We find the possibility of improved constraints when weighting the potential host galaxies by stellar mass and star-formation rate.
Including all galaxies brighter than 0.01 $L_B^\star$ (including 99\% of the total blue luminosity) we find $H_0= 76^{+48}_{-23}$ km s$^{-1}$ Mpc$^{-1}$, or $H_0= 76^{+45}_{-21}$ km s$^{-1}$ Mpc$^{-1}$ when applying B-band luminosity weights (a proxy for star-formation rate). Weighting all galaxies brighter than 0.005 $L_K^\star$ by their K-band luminosity (a proxy for stellar mass), we find $H_0 = 76^{+41}_{-19}$ km s$^{-1}$ Mpc$^{-1}$.

Although this statistical standard siren measurement of $H_0$ is less precise than the counterpart measurement of $H_0 = 76^{+19}_{-13}$ km s$^{-1}$ Mpc$^{-1}$ (for a flat prior and utilizing the distance ansatz in the 3-dimensional sky map; see \S\ref{sec:results}), it nonetheless shows that interesting constraints on cosmological parameters are possible from gravitational-wave sources even in the absence of an optical counterpart and an identification of the unique host galaxy \citep{Schutz,DelPozzo:2012,Chen:H0, MDC}. Although detailed studies find that the measurement of cosmological parameters from the counterpart approach is likely to surpass the statistical approach~\citep{Chen:H0}, the statistical approach offers an important cross-validation of the counterpart standard siren measurements. Furthermore, the statistical approach holds particular promise for binary black hole sources, which are detected at higher rates than binary neutron star systems and are expected to lack electromagnetic counterparts. The inferior quality of the individual $H_0$ measurements for binary black holes (because of the larger localization volumes) may be compensated for by the improved quantity due to the higher detection rates. The binary black holes will also be detected at much greater distances, and in addition to measuring $H_0$ may constrain additional cosmological parameters such as the equation of state of the dark energy.

\acknowledgments
We thank Chihway Chang, Alison Coil and Risa Wechsler for valuable discussions about galaxy properties.
MF was supported by the NSF Graduate Research Fellowship Program under grant DGE-1746045. MF and DEH were supported by NSF grant PHY-1708081. They were also supported by the Kavli Institute for Cosmological Physics at the University of Chicago through NSF grant PHY-1125897 and an endowment from the Kavli Foundation. DEH also gratefully acknowledges support from a Marion and Stuart Rice Award. The authors acknowledge the Italian Istituto Nazionale di Fisica Nucleare (INFN), the French Centre National de la Recherche Scientifique (CNRS) and the Foundation for Fundamental Research on Matter supported by the Netherlands Organisation for Scientific Research, for the construction and operation of the Virgo detector and the creation and support of the EGO consortium.
We are grateful to the LIGO and Virgo collaborations for releasing sky map data for GW170817 at \url{dcc.ligo.org/LIGO- P1800061/public}. \rr{This work has made use of CosmoHub. CosmoHub has been developed by the Port d'Informació Científica (PIC), maintained through a collaboration of the Institut de Física d'Altes Energies (IFAE) and the Centro de Investigaciones Energéticas, Medioambientales y Tecnológicas (CIEMAT), and was partially funded by the "Plan Estatal de Investigación Científica y Técnica y de Innovación" program of the Spanish government. We also acknowledge the First Two Years data release (https://www.ligo.org/scientists/first2years/).}

\appendix
\section{Statistical $H_0$ likelihood}
In this appendix we derive the $H_0$ posterior probability distribution function from Equation~\ref{eq:posterior}.
We write the likelihood for GW and EM data, $x_{\rm GW}$ and $x_{\rm EM}$, given a value of $H_0$ as:
\begin{equation}
\label{eq:likelihood}
p(x_{\rm GW}, x_{\rm EM} | H_0) = \frac{\int p(x_{\rm GW}, x_{\rm EM}, D_L, \Omega, z | H_0) \,dD_L \,d\Omega \,dz}{\beta(H_0)},
\end{equation}
and factor the numerator as:
\begin{equation}
\label{eq:likelihoodfactor}
\begin{split}
\int p(x_{\rm GW}, x_{\rm EM}, D_L, \Omega, z | H_0) \,dD_L \,d\Omega \,dz &= \int p(x_{\rm GW} | D_L, \Omega) p(x_{\rm EM}| z, \Omega) p(D_L | z, H_0) p_0(z, \Omega) \,dD_L \,d\Omega \,dz \\
&= \int p(x_{\rm GW} | D_L, \Omega) p(x_{\rm EM}| z, \Omega) \delta(D_L - \hat{D}_L( z, H_0)) p_0(z, \Omega) \,dD_L \,d\Omega \,dz \\
&= \int p(x_{\rm GW} | \hat{D}_L(z,H_0), \Omega) p(x_{\rm EM}| z, \Omega) p_0(z, \Omega) \,d\Omega \,dz.
\end{split}
\end{equation}

The $H_0$ posterior is related to the likelihood in Equation~\ref{eq:likelihoodfactor} by a prior:
\begin{equation}
p(H_0 \mid x_{\rm GW}, x_{\rm EM}) = p_0(H_0)p(x_{\rm GW}, x_{\rm EM} | H_0).
\end{equation}
This equation is identical to Equation~\ref{eq:posterior} in the main text.
The normalization term $\beta(H_0)$ is given by \citep[see][]{Mandel:2016,Chen:H0}:
\begin{equation}
\label{eq:beta}
\begin{split}
\beta(H_0) &= \int_{x_{\rm GW}>\rm thresh}\int_{x_{\rm EM}>\rm thresh}\int p(x_{\rm GW}, x_{\rm EM}, D_L, \Omega, z | H_0) \,dD_L \,dz \,d\Omega \,d x_{\rm GW}\,d x_{\rm EM} \\
&= \int P^{\rm GW}_{\rm det}(\hat{D}_L(z,H_0),\Omega, z)P^{\rm EM}_{\rm det}(z,\Omega)p_0(z,\Omega) \,d\Omega \,dz \\
&= \int_0^{z_{\rm h}}\int\int P^{\rm GW}_{\rm det}(\hat{D}_L(z,H_0),\Omega, z)p_0(z,\Omega) \,d\Omega \,dz,
\end{split}
\end{equation}
where we assume that only data that is above some threshold is detected, and we define:
\begin{equation}
P^{\rm GW}_{\rm det}(D_L,\Omega, z) \equiv \int_{d_{\rm GW}> \rm thresh}p(d_{\rm GW}|D_L,\Omega, z)\,dd_{\rm GW},
\end{equation}
and similarly:
\begin{equation}
\begin{split}
P^{\rm EM}_{\rm det}(z, \Omega) &\equiv \int_{d_{\rm EM}>\rm thresh}p(d_{\rm EM}|z, \Omega) dd_{\rm EM} \\
&= \mathcal{H}(z_\mathrm{h}-z),
\end{split}
\end{equation}
where $\mathcal{H}$ is the Heaviside step function.
We assume that the EM likelihood is constant with redshift up to a maximum (horizon) redshift, beyond which we assume there are no detectable host galaxies. \maya{In the statistical analysis in which the EM likelihood is assumed to be uninformative, $z_h$ is equivalent to the maximum redshift of our galaxy catalog, or $z_\mathrm{max}$ defined before Equation~\ref{eq:pmissing}.}
We calculate $P^{\rm GW}_{\rm det}$ by assuming a network signal-to-noise ratio threshold of 12 for detection, a monochromatic BNS mass distribution of $1.4$--$1.4 \ M_\odot$, zero spins, and isotropic inclination angles.

In practice, for nearby BNS sources, the term $\beta(H_0)$ is insensitive to the precise details of this calculation or to the choice of $z_\mathrm{h} \gtrsim 0.2$, and is essentially $\beta(H_0) \sim H_0^3$. This can be seen as follows. In LIGO-Virgo's second observing run, detectable BNS sources were within $\sim 100$\,Mpc~\citep{ObservingScenarios}. For $H_0$ values within our prior range, this corresponds to redshifts $z \lesssim 0.07$, which is much smaller than the maximum detectable galaxy redshift, and so we can work in the limit $z_{\rm h} \rightarrow \infty$. We furthermore assume that the large-scale distribution of galaxies is uniform in comoving volume and we use the low-redshift, linear approximation $H_0 = cz / \ D_L$. At the low redshifts of detected BNS events, the redshifting of the GW signal in the detectors is negligible, and so we assume that $P_{\rm det}^{\rm GW}$ depends only on $D_L$ and $\Omega$, and is independent of $z$. With these approximations, we apply a different chain rule factorization to Equation~\ref{eq:likelihoodfactor} and write:
\begin{equation}
p(x_{\rm GW}, dx_{\rm EM} | H_0)\alpha(H_0) = \int p(x_{GW} | D_L, \Omega) p_0(D_L, \Omega) p(x_{EM}| \hat{z}(D_L,H_0), \Omega) \,dD_L \,d\Omega,
\end{equation}
where $\alpha(H_0)$ is a normalization term analogous to $\beta(H_0)$.
With this factorization, we can follow the steps in Equation~\ref{eq:beta} to write $\alpha(H_0)$ as:
\begin{equation}
\alpha(H_0) = \int P_{\rm det}^{\rm GW}(D_L, \Omega)p_0(D_L,\Omega)P_{\rm det}^{\rm EM}(\hat{z}(D_L,H_0),\Omega)\,d\Omega \,dD_L,
\end{equation}
but this is now a constant (independent of $H_0$) because $P_{\rm det}^{\rm EM}(z,\Omega) = 1$.
We can then do a change of variables $dD_L = {c} / {H_0}\,dz$, and if we assume that $p_0(D_L, \Omega) \propto D_L^2$, we get:
\begin{equation}
p(x_{\rm GW}, x_{\rm EM} | H_0) \propto \frac{1}{H_0^3}\int p(x_{\rm GW} | \hat{D}_L(z,H_0), \Omega) p(x_{\rm EM}| z, \Omega) p_0(z, \Omega) \,d\Omega \,dz.
\end{equation}
(Here we have dropped $\alpha(H_0)$ because it is a constant.) This is equivalent to Equation~\ref{eq:likelihoodfactor} if we set $\beta(H_0) \propto H_0^3$.

\section{GW170817-like events}
\begin{figure}
\centering
\includegraphics[width=0.7\textwidth]{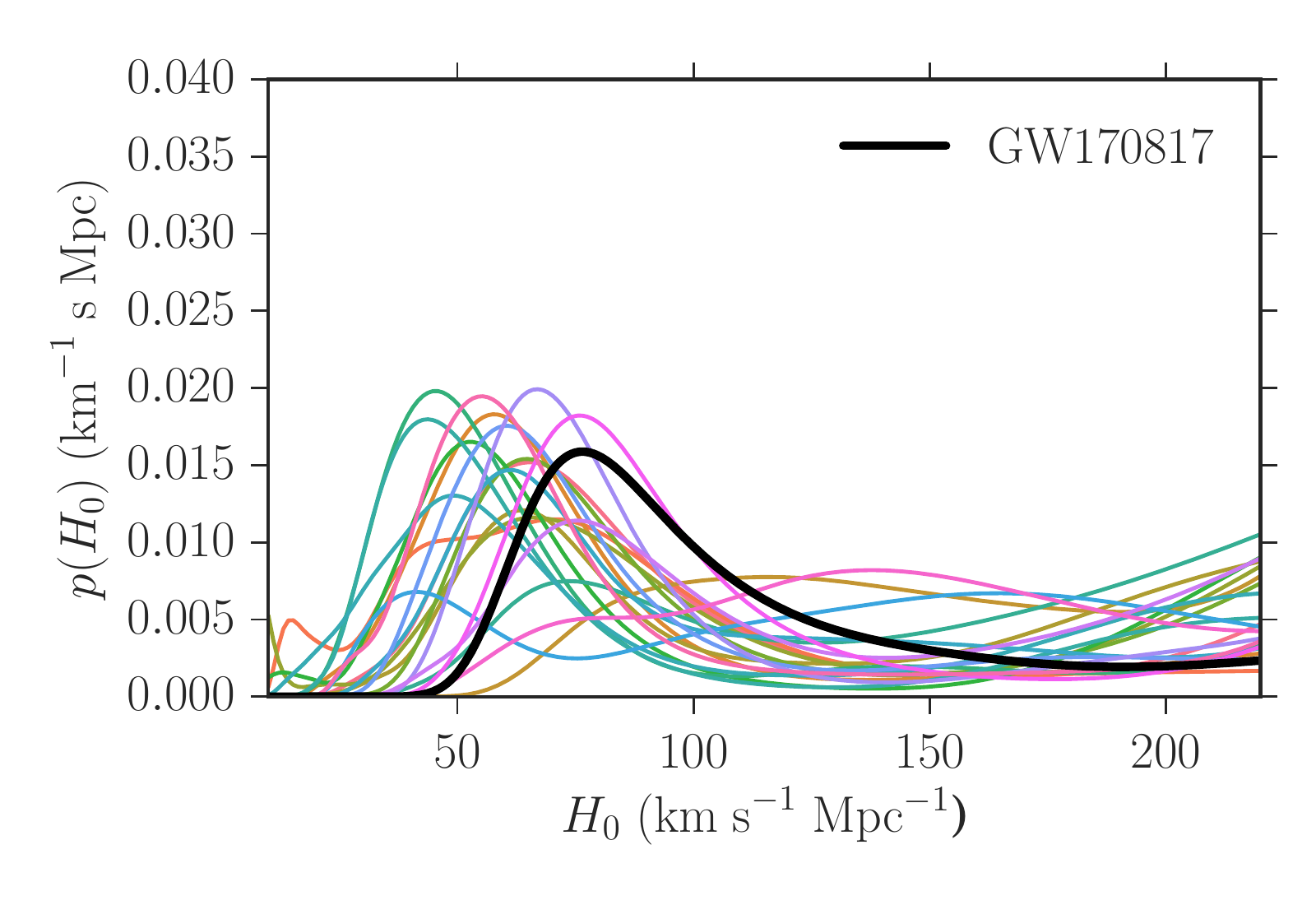}
\caption{\label{fig:GW170817like} $H_0$ posteriors for 20 realizations of the GW170817 3-dimensional sky map centered on different galaxies in the MICE simulated galaxy catalog, assuming $H_0 = 70$ km s$^{-1}$ Mpc$^{-1}$, and incorporating realistic (uncorrected) peculiar velocities with 1-$\sigma$ uncertainties of 400 km/s. The real $H_0$ posterior using the GLADE galaxy catalog is shown in black. It is a typical result for a source with such a small localization volume, as such sources tend to produce a single major peak in the $H_0$ posterior.}
\end{figure}

In order to explore whether the large-scale structure in the GW170817 localization volume, and the resulting statistical $H_0$ posterior, is typical for galaxies at $z \sim 0.01$, we rotate the true GW170817 sky map to different galaxies in the the MICE simulated catalog and repeat the statistical $H_0$ measurement. We assume that unlike for real galaxies in the GW170817 localization volume, no detailed observations have been carried out to measure the peculiar velocity field and apply group corrections. We therefore use the un-corrected redshifts given in the MICE catalog, which include a peculiar velocity contribution. The distribution of peculiar velocities is approximately described by a Gaussian of width 400 km/s, and we incorporate this uncertainty in the simulated $H_0$ measurements. Figure~\ref{fig:GW170817like} shows the results for 20 realizations of the GW170817 localization volume centered on different galaxies in the MICE catalog. We see that the true GW170817 statistical $H_0$ measurement (we once again use the K-band luminosity-weighted $L_K > 0.1 L_K^\star$ posterior shown in the right panel of Figure~\ref{fig:H0post_lumweights} as a representative posterior) is fairly typical among the different realizations. Over 50 different realizations, the information, given by the difference in Shannon entropy between prior and posterior, is $0.43^{+0.43}_{-0.19}$ bits (median and symmetric 90\% intervals), whereas the information for the true GW170817 measurement is 0.67 bits. If we lower the peculiar velocity uncertainty in the simulations from 400 km/s to 200 km/s, the GW1708170-like posteriors become slightly more informative on average, with a typical information gain of $0.57^{+0.42}_{-0.27}$ bits.

\bibliographystyle{yahapj}
\bibliography{references}



\end{document}